\documentclass[aps, pra, superscriptaddress, reprint, twocolumn,floatfix]{revtex4-1}
\usepackage[usenames,dvipsnames]{xcolor}
\usepackage{appendix}
\usepackage{ulem}
\usepackage{amssymb}
\usepackage{graphicx}
\usepackage{amsmath}
\usepackage{hyperref}
\hypersetup{
 colorlinks=true,
 linkcolor=blue,
 anchorcolor = blue,
 citecolor = blue,
 filecolor = blue,
 urlcolor = blue
}
\usepackage{braket}
\usepackage{subcaption}
\usepackage{bbm}
\usepackage{dcolumn}
\usepackage{braket}
\usepackage{bm}
\usepackage{amsmath,amsthm,amssymb}
\usepackage{color}
\usepackage{verbatim}
\usepackage{comment}
\usepackage{dsfont}
\usepackage{physics}
\usepackage[justification=Justified]{caption}
\usepackage{float}
\usepackage{babel}

\begin{document}
\newfloat{suppfig}{tbh}{losf}
\floatname{suppfig}{SM Figure }
\setcounter{suppfig}{0}

\title{Observing the quantum Mpemba effect in quantum simulations}

\newcommand{\IQOQI}{\affiliation{Institute for Quantum Optics and Quantum Information, Austrian Academy of Sciences, Technikerstra{\ss}e 21a, 6020 Innsbruck, Austria}}
\newcommand{\UIBK}{\affiliation{University of Innsbruck, Institute for Experimental Physics,  Technikerstra{\ss}e 25, 6020 Innsbruck, Austria}}
\newcommand{\AQT}{\affiliation{AQT, Technikerstra{\ss}e 17, 6020 Innsbruck, Austria}}
\newcommand{\ITP}{\affiliation{University of Innsbruck, Institute for Theoretical Physics, Technikerstra{\ss}e 21a, 6020 Innsbruck, Austria}}
\newcommand{\SISSA}{\affiliation{SISSA and INFN, via Bonomea 265, 34136 Trieste, Italy }}
\newcommand{\ICTP}{\affiliation{International Centre for Theoretical Physics (ICTP), Strada Costiera 11, 34151 Trieste, Italy}}
\newcommand{\Grnbl}{
\affiliation{Univ.\  Grenoble Alpes, CNRS, LPMMC, 38000 Grenoble, France}}
\newcommand{\Caltech}{\affiliation{Walter Burke Institute for Theoretical Physics, and Department of Physics and IQIM, Caltech, Pasadena, CA 91125, USA}}
\newcommand{\s}{\mathbf{s}}
\renewcommand{\tr}{\mathrm{Tr}}

\author{Lata Kh Joshi}
\IQOQI
\ITP
\SISSA

\author{Johannes Franke}
\IQOQI
\UIBK

\author{Aniket Rath}
\Grnbl

\author{Filiberto Ares}
\SISSA

\author{Sara Murciano}
\Caltech

\author{Florian Kranzl}
\IQOQI
\UIBK

\author{Rainer Blatt}
\IQOQI
\UIBK

\author{Peter Zoller}
\IQOQI
\ITP

\author{Beno\^it Vermersch}
\IQOQI
\ITP
\Grnbl

\author{Pasquale Calabrese}
\SISSA
\ICTP

\author{Christian F. Roos}
\IQOQI
\UIBK

\author{Manoj K. Joshi}\email{manoj.joshi@uibk.ac.at}
\IQOQI
\UIBK

\begin{abstract}
The non-equilibrium physics of many-body quantum systems harbors various unconventional phenomena. In this letter, we experimentally investigate one of the most puzzling of these phenomena— the quantum Mpemba effect, where a tilted ferromagnet restores its symmetry more rapidly when it is farther from the symmetric state compared to when it is closer.
We present the first experimental evidence of the occurrence of this effect in a trapped-ion quantum simulator. The symmetry breaking and restoration are monitored through entanglement asymmetry, probed via randomized measurements, and post-processed using the classical shadows technique. Our findings are further substantiated by measuring the Frobenius distance between the experimental state and the stationary thermal symmetric theoretical state, offering direct evidence of subsystem thermalization.

\end{abstract}
\maketitle

\paragraph*{Introduction--}When a system is brought out of equilibrium, it may exhibit phenomena that defy conventional wisdom. One particularly puzzling example is the Mpemba effect, first described as the phenomenon where hot water freezes faster than cold water \cite{Mpemba1970}, and then extended to a wide variety of systems \cite{lasantaMpemba,lu2017nonequilibrium, klich2019mpemba, kumar2020exponentially, Bechhoefer2021, kumar2022anomalous,Teza2023a}. An anomalous relaxation, reminiscent of the Mpemba effect, can manifest at zero temperature in isolated many-body quantum systems. Specifically, starting from a configuration that breaks a symmetry, its restoration can happen more rapidly when the initial state shows a greater degree of symmetry breaking \cite{Ares2023}. This behavior, dubbed as \textit{quantum Mpemba effect} (QMPE), is driven by entanglement and quantum fluctuations. The origin and ubiquity of the QMPE are active areas of research. For instance, in integrable systems, the conditions under which the QMPE occurs are now well understood~\cite{Rylands2023}. However, the existence of this effect remains outstanding in generic quantum systems— such as non-integrable and synthetic quantum many-body systems. In this regard, we experimentally explore the QMPE in a chain of spins coupled via power-law decaying interactions. 

Present-day programmable quantum simulators, with their impeccable ability to create, manipulate, and analyze quantum states, provide us with excellent test beds to examine non-equilibrium dynamics in many-body quantum systems~\cite{Zhang2017, Joshi2022, Scholl2021, Britton2012}. To address our objective of investigating the QMPE, we employ a trapped-ion quantum simulator consisting of $N=12$ interacting spin-$1/2$ particles. The system is initialized into a product state where each spin points in the $z$ direction and subsequently all spins are tilted by an angle $\theta$ from the $z$ axis. 
The spin states designated by $\theta=0 $ and $\pi$ (we will refer to them as ferromagnetic states) are $U(1)$ symmetric as they remain invariant under a rotation about the $z$ axis; conversely for $0 < \theta < \pi$ (tilted ferromagnets), the states explicitly break such symmetry. The tilted ferromagnetic state is then evolved with the engineered XY Hamiltonian \cite{Porras2004, Jurcevic2014, Monroe2021} that also possesses the $U(1)$ symmetry and determines the properties of time-evolved states. The $U(1)$ symmetry implies the conservation of the magnetization along the $z$ axis, i.e. the conserved charge is defined as $Q=1/2\sum_j \sigma_j^z$. In the thermodynamic limit, the reduced density matrix of a subsystem relaxes to a Gibbs ensemble~\cite{Deutsch1991, Srednicki1994, rigol2008}, and the initially broken symmetry is restored~\cite{Ares2023}. The latter is a direct consequence of the Mermin-Wagner theorem which forbids spontaneous breaking of a continuous symmetry in one-dimensional systems at finite temperatures \cite{Hohenberg1967,Mermin1966}.  

\begin{figure*}[t!]
\includegraphics[width=\linewidth]{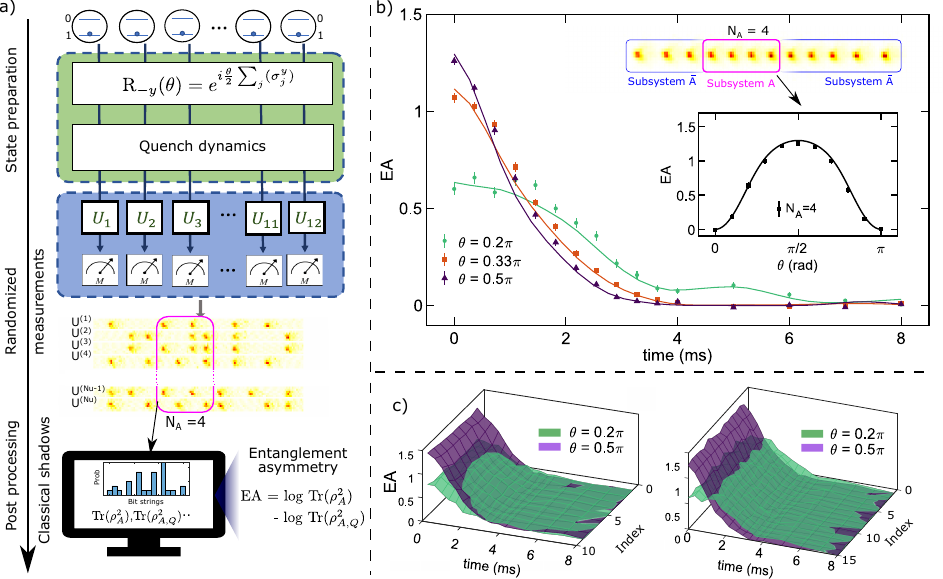}
\caption{a) Protocol to measure the EA  with a 1d array of  12 trapped ions using randomized measurements. After rotating the initially prepared ferromagnetic state by an angle $\theta$, the system is evolved (quenched) with the Hamiltonian of interest for time $t$. Subsequently, randomized local rotations and projective measurements are performed. The obtained bitstrings (constructed from site-resolved fluorescence images) are then classically processed to estimate the EA. Under the quench dynamics with XY Hamiltonian~\eqref{eq:IsingH}: b) Experimentally measured EA, averaged over all possible subsystems of size $N_A =4$ chosen from ion indices between 4 and 9, is plotted for angles $\theta = 0.2\pi$, $0.33\pi$ and $0.5\pi$. A fast restoration of symmetry is observed for the largest angle, confirming the quantum Mpemba effect (QMPE). Inset presents the EA at initial time $t=0$ as a function of the tilt angle $\theta$.  Solid curves denote theory including  decoherences and symbols are experimental data. Error bars are calculated via the jackknife resampling method (see SM section H). c) The measured EA is plotted for all connected 4-qubit subsystems of the whole chain (on the left)  and for all possible 4-qubit subsystems (connected and disconnected) between ion numbers 4 to 9 (on the right). These plots show the independence of the QMPE on the subsystem. The $x$ axis represents the subsystem index detailed in the main text.}
\label{fig:fig1}
\end{figure*}

To set up an experimental indicator of symmetry-breaking, we consider a bipartition of the system as $A\cup \bar A$. The charge $Q$ that generates the $U(1)$ symmetry decomposes into the contribution of each subsystem, i.e. $Q=Q_A+Q_{\bar{A}}$. The possible values $q$ of the subsystem charge $Q_A$ define the charge sectors. When a reduced density matrix is symmetric,  it is block diagonal in the eigenbasis of $Q_A$. Based on this property, the restoration of the symmetry, at the level of a subsystem $A$, is estimated by the ``Entanglement Asymmetry" (EA), defined as \cite{Ares2023},
\begin{equation}
\mathrm{EA}  = \log[\tr(\rho_A^2)] -\log[\tr(\rho_{A,Q}^2)],~
\label{eq:EA}
\end{equation}
where $\rho_{A,Q}$ denotes the symmetrized counterpart of $\rho_A$  i.e. $\rho_{A,Q} = \sum_{q \in  \mathbb{Z}} \Pi_q \rho_{A} \Pi_q $. Each $\Pi_q$ is the projector onto the charge sector $q$ in the subsystem $A$. The EA is a non-negative quantity that vanishes if and only if $\rho_A$ is symmetric, that is $\rho_A=\rho_{A, Q}$ \cite{Ares2023}. 
The EA has proven to be an effective tool for investigating broken symmetries, not only in out-of-equilibrium many-body systems~\cite{Rylands2023, khor2023confinement} but also in quantum field theories~\cite{capizzi2023entanglement,chen2023entanglement, capizzi2023universal} and black hole physics~\cite{ares2023page}. For present studies, both parts of the above equation can be estimated using the classical shadow formalism from randomized measurements~\cite{elben_review, Huang2020}.

In the current context, the QMPE manifests when an input state with a larger value of the EA, i.e. a greater degree of symmetry breaking, undergoes faster symmetry restoration than the one with a smaller value of the EA. Studying EA as a function of time, the QMPE  is identified by the presence of a crossing between the EA curves for two states which are initialized at different degrees of symmetry-breaking \cite{Ares2023}. This route to symmetry restoration can also be mapped via a state distance, such as the Frobenius distance~\cite{fagotti2013reduced}, which will be presented as a complementary approach to studying the QMPE. These studies performed on a quantum simulator not only allow us to investigate the QMPE in a non-integrable system but also provide insights into the robustness of the effect under realistic physical conditions. In contrast to other quantum versions of the Mpemba effect, which require an external reservoir to drive the system out of equilibrium \cite{Nava2019,cll-21,manikandan-21,kck-22,ias-23, Chatterjee2023, chatterjee2023multiple}, our studies involve an isolated quantum system undergoing a unitary evolution of a pure quantum state. We note that during the preparation of this letter, the Mpemba effect with a single ion coupled to an external reservoir has been reported~\cite{Shahaf2024, zhang2024}.  These investigations complement our work while examining the QMPE in a different scenario.

Our experiment marks the first observation of the QMPE in a many-body quantum system. We show that the presence of integrability-breaking interactions and decoherence do not undermine its occurrence. This Letter presents experimental studies of symmetry restoration of a tilted ferromagnet for various scenarios. At first, we present our findings for nearly unitary dynamics of a tilted ferromagnet; the experimental results show that the QMPE occurs for the interacting XY spin chain.  We also explore the QMPE under local disorders added to the interacting spin chain, and the scenario where tilted spins solely interact with the environmental noise. The former case reveals the QMPE for weak disorder strengths, however, when the disorder reaches a certain strength, it slows down the symmetry restoration, and subsequently weakens the presence of the QMPE. In the latter case, we do not find the QMPE.

\paragraph*{Setup--}We use a $N=12$ qubit trapped-ion quantum simulator to study the QMPE. A linear string of calcium ions is held in a Paul trap and laser-cooled to the motional ground state. The spin states are encoded into two long-lived electronic states, $\ket{S_{1/2}, m=1/2} \equiv \ket{1} \equiv \ket{\downarrow}$ and $\ket{D_{5/2}, m=5/2} \equiv \ket{0} \equiv \ket{\uparrow}$ and manipulated by a narrow-linewidth 729~nm laser. Further experimental details are given in the Supplemental Material (SM) sections A, B, C, and D~\cite{LKJSM2024}\nocite{Arrazola2016, cieslinski2023analysing,Haah2017,Brydges2019, Rath2021, Satzinger2021, Zhou2020, Elben2020b, Neven2021, Imai2021,zhang2023experimental, Rath_QFI_2021, vitale2023estimation, LKJ_2022_chaos, Mezzadri2006, Hoeffding1992, Vitale2022, Rath2023OE, wu1986jackknife}. The experimental recipe for investigating the QMPE using randomized measurements is presented in Fig.~\ref{fig:fig1}a. The spin state is initialized in a ferromagnetic product state \mbox{$\ket{\downarrow}^{\otimes N}$} and subsequently all spins are tilted by an angle $\theta$ using a laser beam that resonantly couples the two qubit states. The tilted state is then time-evolved (quenched) with the desired Hamiltonian.
After performing local random rotations $U$, site-resolved projective measurements are performed on the time-evolved state. The observables related to symmetry-breaking and the state distance, i.e. the EA and the Frobenius distance for the subsystem of interest $A$, are estimated from the classical shadows. In sections E, F and G of SM~\cite{LKJSM2024} we provide details of the estimators.

\paragraph*{Symmetry restoration with a long-range spin-spin interaction--}The experimental data showing the QMPE are presented in Fig.~\ref{fig:fig1}b. We choose three tilt angles: $\theta=0.2\pi$, $\theta=0.33\pi$, and $\theta=0.5\pi$ in increasing order of symmetry-breaking at $t=0$, i.e. EA = 0.64, 1.12 and 1.3 for $N_A =4$, respectively. The tilt angle dependence of the EA at initial time $t=0$ is presented in the inset of Fig.~\ref{fig:fig1}b for subsystem size $N_A =4$. The tilted ferromagnetic states are then quenched with a $U(1)$ symmetric Hamiltonian engineered in our experiment. We engineer a power-law decaying XY interaction expressed as the Hamiltonian, 
\begin{equation}
H_{\mathrm{XY}} = \sum_{i>j}\frac{J_0}{2|i-j|^\alpha} \left(\sigma^x_i \sigma^x_j + \sigma^y_i \sigma^y_j\right)~,
\label{eq:IsingH}
\end{equation}
where $\sigma^a_i$ denote the Pauli matrices for $a=x,y$ at lattice site $i=1\dots N$. The realized interaction strength and range are $J_0 \approx 560~\mathrm{rad/s}$ and  $\alpha \approx 1$, respectively. 

Under the time evolution with the XY Hamiltonian, in the thermodynamic limit, the subsystem initialized into an asymmetric state is expected to attain the same symmetry of the Hamiltonian, i.e. the subsystem attains U(1) symmetry \cite{Ares2023}. The symmetry restoration for subsystems of size $N_A=4$ is monitored by measuring the EA at various times; see Fig.~\ref{fig:fig1}b. The plotted EA is the average over all possible subsystems with size $N_A=4$ from the central 6 lattice sites. We witness that the EA decreases and tends to zero for the three tilt angles, indicating that the subsystem state $\rho_A$ attains $U(1)$ symmetry. The striking feature is that the EA decays faster for a state at $\theta=0.5\pi$ than for states at $\theta=0.33 \pi$ and $ 0.2 \pi$; i.e., in Fig.~\ref{fig:fig1}b the purple curve reaches zero before the green or the orange curve. Naively, one might assume that the state with the smallest initial EA would restore the symmetry at the earliest. The observed crossing in the  EA curves confirms the QMPE. 
In Fig.~\ref{fig:fig1}b, the solid lines are numerical simulations carried out for the experimental conditions, while including decoherence effects discussed in SM~\cite{LKJSM2024}. Notably, our system $N=12$ sufficiently captures the features of the QMPE. Decreasing the system size results in a finite size effect and affects the observation (see SM section I~\cite{LKJSM2024}). On the other hand, increasing the system does not substantiate the observation but increases the experimental complexity.

We further examine the robustness of the QMPE with respect to the choice of subsystem $A$. For this, we evaluate the EA for various subsystems with fixed size $N_A=4$ and plot the experimental results in Fig.~\ref{fig:fig1}c for tilt angles $\theta =0.2\pi$ (green surface)  and $\theta =0.5\pi$ (purple surface). In the left panel, we display the results for subsystems that are connected. We have considered all connected subsystems of size $N_A=4$ from one edge of the chain to the other edge. There are $9$ of them which are $[1,2,3,4],[2,3,4,5],[3,4,5,6], \cdots$, here the integers denote ion indices. Notably, the subsystems that lie at the edges of the ion string display the crossing of the EA curves at later times than those from the middle of the chain (bulk region). This is attributed to the effect of boundaries, as the subsystems from the edge regions are accompanied by a lesser number of neighboring particles than the ones in the bulk. This effect becomes more pronounced for smaller system sizes, which we further discuss in SM~\cite{LKJSM2024}. Furthermore, the choice of the subsystem is extensively studied by considering all connected and disconnected subsystems from the bulk. In the right panel of Fig.~\ref{fig:fig1}c, we display the EA for the two tilt angles as a function of time and subsystem index. Here, we consider all subsystems of size $N_A=4$ forming 15 subsystems out of the central $6$ ions. In contrast to the left panel of Fig.~\ref{fig:fig1}c, the crossing time of the EA for the two tilt angles does not depend upon the choice of the subsystem, which implies that the QMPE in the bulk region is robust against the choice of the subsystem.

\begin{figure}[t!]
\includegraphics[width=1\linewidth]{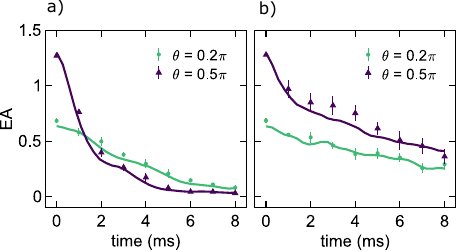}
\caption{Time evolution of the EA with the disorder Hamiltonian \eqref{eq:disorderH}. a) In the presence of weak disorders,  $h_i\in 6(0, J_0)$, we observe crossing in the EA for the two tilt angles as a function of interaction time, thus revealing the QMPE in the measurements. b) For the strong disorders, $h_i\in 14(0, J_0)$, the QMPE is not visible in the measured data. The presented data are the averaged EAs for all 4 qubit subsystems out of the central 6 qubits. Solid curves denote theory and symbols are experimental data where the error bars are the standard deviation of the mean over the disorder realizations}. 
\label{fig:disorder}
\end{figure}

\paragraph*{QMPE in the presence of disordered interactions --}
A central theme in modern research is to understand how localization alters the relaxation and thermalization dynamics in spin systems~\cite{abanin2019}. The versatility of our experimental setup enables us to study this interesting problem through the lens of the EA and the QMPE. On the experimental side, we add  transverse disorder terms to our XY Hamiltonian, thus engineering a disorder Hamiltonian of the type,
\begin{equation}
H =\sum_{i>j}\frac{J_0}{2|i-j|^\alpha} \left(\sigma^x_i \sigma^x_j + \sigma^y_j \sigma^y_i\right)+ \sum_i h_i \sigma^z_i~.
\label{eq:disorderH}
\end{equation}
The last term denotes static disorder terms realized in the experiment with site-dependent light shift laser beams~\cite{Maier2019}. The disorder fields are chosen randomly from a uniform distribution with disorder strengths $h_i\in w(0, J_0)$.
The Hamiltonian \eqref{eq:disorderH} is $U(1)$ symmetric for each disorder realization and so can be exploited to study the QMPE.  
The tilted ferromagnetic states for two choices of tilt angles $\theta=0.2\pi$ and $0.5\pi$ are evolved under this Hamiltonian. We consider weak ($w = 6$) and strong ($w = 14$) disorder configurations. The experiment is repeated for 5 disorder sets of the weak and strong disorder cases~(see SM~\cite{LKJSM2024} for the measured disorder values). In Fig.~\ref{fig:disorder}, we show the EA averaged over the 5 disorder realizations and all 15 subsystems of size 4 out of the central 6 ions. For the weak disorder, we observe the QMPE as the EA curves cross for the two tilt angles, indicating a faster drop of the EA for the input state with a larger EA at $t=0$.  These results prove the robustness of the QMPE for weak disorders. Oppositely, for strong disorders, there is no crossing between the EA curves within the experimental time window. In this case, the strong disorders localize the interactions to single sites thus preventing the subsystem thermalization.

\begin{figure}[t!]
\includegraphics[width=0.9\linewidth]{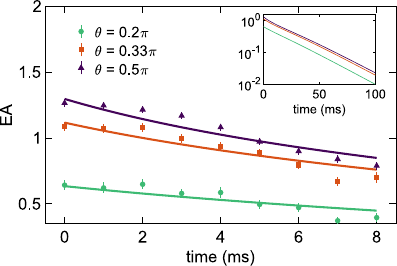}
\caption{ Symmetry restoration under a pure dephasing evolution for three tilted ferromagnetic states; $\theta =0.2\pi, 0.33\pi,$ and $0.5\pi$. Solid lines correspond to numerical simulations and data points are the experimental results averaged over 15 subsystems of $N_A=4$ generated from the central 6 ions and the error bars are calculated through jackknife resampling (see SM section H).  Inset: simulation results are presented for late times. A reduction of the EA for all three tilt angles implies symmetry restoration, however, the QMPE is absent for the pure dephasing case since the EA curves do not cross.
}
\label{fig:puredephasing}
\end{figure}
\paragraph*{Relaxation of tilted ferromagnetic states under dephasing --} 
So far we have studied the QMPE under near-unitary time evolution and investigated the  QMPE for tilted ferromagnets evolved under the XY and XY+disorder Hamiltonian. Now we will explore a different scenario where the tilted ferromagnet evolves under a non-unitary evolution such as pure dephasing of the spins with the environmental noise. We turn the spin-spin interaction off and let the system freely evolve for time $t$. Due to the interaction with the fluctuating magnetic field in the laboratory and the phase noise of the laser beam (see SM~\cite{LKJSM2024}), dephasing takes place in the experiment. Under this scenario, the initially asymmetric state is expected to symmetrize. 

The experimental data are presented in Fig.~\ref{fig:puredephasing} and the results are corroborated by numerical simulations. The measurement protocol is similar to the one presented in Fig.~\ref{fig:fig1}a except that the states are time-evolved under the ambient experimental noise.
In this case, we observe the restoration of symmetry, however, the crossing of the EA for various tilt angles is not observed. The experimental data and numerical simulations, see inset, imply that pure dephasing evolution does not show the QMPE while undergoing symmetry restoration of initially asymmetric states. In Fig.~\ref{fig:puredephasing}, solid lines correspond to numerical simulations, which are carried out for the experimentally measured dephasing rates listed in section C of SM. 

\paragraph*{Frobenius distance to observe QMPE--}
\begin{figure}[t!]
    \includegraphics[width=0.9\linewidth]{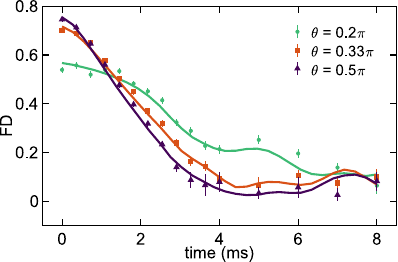}
\caption{The Frobenius distance between the experimental time-evolved state and the theoretical equilibrium state is shown for three tilt angles and subsystem size $N_A = 4$. The state that is initially farthest from its equilibrium state attains the latter at the earliest time; confirming the occurrence of the QMPE in quantum simulations via Frobenius distance measures. Solid lines are numerical simulations and the symbols are experimental data averaged over subsystems from the central region. The error bars are calculated via the jackknife resampling method (see SM section H)}. 
\label{fig:FrobDist}
\end{figure}
In a complementary approach, the route to symmetrization can also be probed by measuring the distance between two states --- the experimental time-evolved state $\rho_A(t)$ and the diagonal ensemble $\rho_{A}^{\rm DE}$~\cite{rigol2007}. The diagonal ensemble describes the average behavior of any observable at long times. It is a mixed state that is diagonal in the eigenbasis of the quenching XY Hamiltonian and, therefore, is $U(1)$ symmetric. According to the eigenstate thermalization hypothesis, for large systems, this ensemble is equivalent to a Gibbs  ensemble~\cite{Deutsch1991, Srednicki1994, rigol2008}.   
From the data obtained in the randomized measurements, we evaluate the Frobenius
distance~\cite{fagotti2013reduced} between $\rho_A(t)$ and $\rho_{A}^{\rm DE}$ (see more details in section G of SM)

The results are presented in Fig.~\ref{fig:FrobDist} for three tilt angles $\theta=0.2\pi, 0.33\pi$ and $0.5\pi$. From the present analysis, we note that the states which start farther away from the corresponding diagonal ensemble attain it earlier than those which start comparably closer. As we see, the state at an angle $0.5\pi$ begins at the largest distance and relaxes to the diagonal ensemble the fastest. Such observation is the manifestation of the QMPE via Frobenius distance measurements. Our findings constitute the first experimental evidence that the subsystem as a whole attains a stationary Gibbs ensemble. Although there exist previous experimental studies on the thermalization of isolated quantum systems~\cite{Gring2012, trotzky2012, kaufman2016, neill2016, ueda2020, zhou2022}, they predominantly focused on specific local observables.

\paragraph*{Conclusions and outlook--}
In this Letter, we have presented the first experimental demonstration of the QMPE. Notably, this phenomenon manifests itself at times much shorter than those relative to finite-size effects, such as revivals. This separation of time scales enables us to observe the QMPE distinctly in our experimental setup, comprising of a modest number of qubits ($N=12$),  even in the presence of decoherence and disorder. Our investigation employs two distinct quantities, the entanglement asymmetry and the Frobenius distance, establishing that the QMPE is not specific to a particular observable. This versatility paves the way for further investigation employing a variety of theoretical and experimental tools.

Our focus in this study has been on a specific class of initial states— the tilted ferromagnets. However, our findings motivate future studies with other configurations to determine the experimental conditions for the occurrence of the QMPE in ergodic systems. While the role of entanglement in driving thermalization in isolated quantum systems is well understood, it will be important to elucidate the interplay between entanglement and quantum fluctuations for the observation of the QMPE in generic quantum systems.
The measurement protocol used in this work is easily applicable to other platforms with individual control and readout capabilities such as arrays of atoms in optical lattices, Rydberg systems, or superconducting qubits~\cite{Daley2022, Bloch2008, Evered2023, Monroe2021, Hoke2023}. This spotlights tantalizing opportunities for the experimental investigation of the QMPE, especially in higher dimensional systems where a richer phenomenology is expected due to the possibility of spontaneous symmetry breaking at finite temperatures. Our experimental findings may stimulate further theoretical investigations about the role played by dissipation and disorder in the dynamical restoration of symmetry. 
We believe that further studies on the QMPE will provide new protocols for faster preparation of thermal states (albeit at the subsystem level). Such states can then be used as input states for quantum simulation experiments. The experimentally accessible quantities discussed in the present manuscript, such as the EA and the Frobenius distance, will be valuable tools to assess the quality of the state preparation.

\paragraph*{Acknowledgments--}
The experimental team (JF, FK, MKJ, CR, RB) would like to acknowledge funding from the Institut f\"ur Quanteninformation GmbH and European Union's Horizon 2020 research and innovation programme under grant agreement No 101113690 (PASQuanS2.1). LKJ, BV and PZ acknowledge funding from the Austrian Science Foundation (FWF, P 32597 N). LKJ acknowledges financial support from the PNRR MUR project PE0000023- NQST, and  HPC and supercomputer facilities of the University of Innsbruck,  where most of the numerical simulations were carried out. PZ acknowledges the funding from the Simons Collaboration on Ultra-Quantum Matter, which is a grant from the Simons Foundation (651440). PC and FA are supported by ERC under Consolidator Grant number 771536 (NEMO).
SM acknowledges the support from the Caltech Institute for Quantum Information and Matter and the Walter Burke Institute for Theoretical Physics at Caltech.
Work in Grenoble is funded by the French National Research Agency via the JCJC project QRand (ANR-20-CE47-0005), and via the research programs Plan France 2030 EPIQ (ANR-22-PETQ-0007),  QUBITAF (ANR-22-PETQ-0004) and HQI (ANR-22-PNCQ-0002).
A.R.\ acknowledges support by Laboratoire d'excellence LANEF in Grenoble (ANR-10-LABX-51-01) and from the Grenoble Nanoscience Foundation. In numerical simulations, we have used the quantum toolbox QuTiP \cite{qutip}.
\section*{Supplemental material}
\subsection{Experimental apparatus}
\label{sec:experimentalsetup}
In this work, we used a trapped-ion quantum simulator capable of storing and manipulating long ion strings for studying the QMPE. The experimental setup consists of a string of $N=12$ $^{40}$Ca$^+$ ions that is made to emulate a spin-$1/2$ chain with long-range interactions. Each qubit is labeled by basis vectors $s = (\ket{\downarrow}, \ket{\uparrow}) \equiv  (\ket{1}, \ket{0})$, which are encoded onto $\ket{\text{S}_{1/2}, m=+1/2}$ and $\ket{\text{D}_{5/2}, m=+5/2}$ electronic levels of a calcium ion. The power-law XY interaction expressed in Eq.~2 (see main manuscript) is engineered with the help of a global laser beam that interacts with ions' electronic and motional degrees of freedom from the transverse direction of the ion chain \cite{Jurcevic2014, Porras2004}. Furthermore, the disordered Hamiltonian displayed in Eq.~3 (see main manuscript) is generated by shining a set of site-selective light shift beams along with the global entangling beam \cite{Maier2019}. In the experimental setup, we have noticed that the time-evolved states experience dephasing while undergoing the time evolution. The dephasing rates have been measured for all cases presented in this letter. The coherence times, which are related to dephasing rates via $\Gamma = 1/(2T_{\mathrm{coh}})$, are $T_{\mathrm{coh}}\sim 16$~ms for the XY interaction case,  $T_\mathrm{coh}\sim 45$~ms for the pure dephasing cases, $T_{\mathrm{coh}}\sim27(2)$~ms for the strong disorders and $T_{\mathrm{coh}}\sim27(4)$~ms for the weak disorder cases. 

\subsection{AC-Stark shift compensation during the spin-spin interaction} 
\label{sec:XYHamiltonian}
The trapped-ion platforms employ an Ising interaction when a bichromatic laser beam off-resonantly excites the motional red and blue sidebands of an ion chain \cite{Porras2004}. In the case of symmetric detuning $\Delta$ of the bichromatic beam from the motional sidebands, an AC-Stark shift is present due to the coupling of the laser to the ion's spectator levels. For example, in our setup, when the interaction is generated on the quadrupole transition of a calcium ion chain with a 729-nm laser beam, the off-resonant coupling to the dipole-allowed transitions gives an AC-Stark shift. We cancel this shift by adding a third frequency component to the laser beam, such that the light off-resonantly couples to the quadrupole transition and induces an AC-Stark shift that is of opposite sign but of equal strength compared to the dipole transitions. 

The Ising interaction is then transformed into an XY interaction via asymmetrically driving the sidebands by $\Delta \pm \delta_c$. In the limit of $\delta_c$ being smaller than $\Delta$ and much larger than the spin-spin coupling strength $J_0$, the interaction Hamiltonian is written as \cite{Arrazola2016},
\begin{equation}
\label{eq:XYwithACShift}
    H=\sum_{i>j}\frac{J_0}{2|i-j|^\alpha} \left(\sigma^x_i \sigma^x_j + \sigma^y_j \sigma^y_i\right)+ \frac{\delta_c}{2\Delta}\sum_i B_i \sigma^z_i~,
\end{equation}
where an additional AC-Stark shift proportional to $\delta_c$ is induced. This term is often irrelevant when the input states are initialized in the eigenbasis of the $\sigma_z$ operator. However, it is relevant when the input state is not the eigenstate of the $\sigma_z$ operator, which is currently the case for $\theta \ne 0$. In the above equation, $B_i$ depends upon the motional occupation number $\langle N_\ell \rangle$, site-dependent Rabi frequency $\Omega_i$ and mode dependent Lamb-Dicke parameter $\eta_{i,\ell}$; and it is expressed as 
\begin{equation}
    B_i= \Omega_i^2 \Delta \sum_\ell \frac{\eta_{i,\ell}^2}{\Delta_\ell^2}(\langle N_\ell \rangle + 1/2). 
\end{equation}
Here, $\Delta_\ell$ is the laser detuning from the $\ell^{th}$ motional sideband. The shift can be different for each ion due to its dependence on the Lamb-Dicke parameter which is different for each ion. However, the major contribution comes from the center of mass mode, which is closest to the laser drive frequency, and it is eliminated by appropriately adjusting the intensity of the third frequency component of the laser beam. 

\subsection{Experimental noise}
\begin{suppfig*}
    \centering
 \includegraphics[width=1\linewidth]{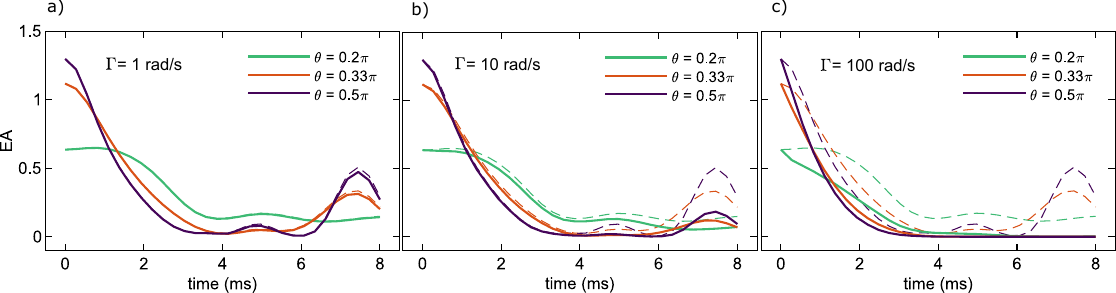}
    \caption{ We numerically examine the effect of global dephasing on EA and on the observation of the QMPE.  Dashed lines correspond to unitary dynamics and solid lines correspond to dephasing dynamics. With increasing dephasing rates; a) $\Gamma =1$ rad/s, b) $\Gamma =10$ rad/s, and c) $\Gamma =100$ rad/s, a reduction in EA is observed for all three tilt angles. Conversely, the QMPE feature i.e. crossing of EA curves for three tilt angles is present even for the highest dephasing rate.} 
    \label{fig:cohLoss}
\end{suppfig*}
\label{sec:dephasing}
The dominant decoherence effect in our experiment is that of dephasing noise, an unwanted noise that randomly rotates the spin vector along the $z$ axis. In our system, the dephasing is a combination of laser phase noise/frequency noise and magnetic-field noise. 
We measure the overall effect of noise by performing Ramsey measurements for entangling interaction, disorder, and pure dephasing cases. For the XY quenches, the measured coherence time is $T_\mathrm{coh}\sim 16$~ms. The coherence time here is measured such that we could account for laser phase noise, magnetic field noise, and fluctuating AC-Stark shift due to thermal occupation of the motional modes. Specifically, the measured coherence time without turning on the quench beam is $T_\mathrm{coh}\sim 45$~ms.

A shorter coherence time in the case of entangling interactions can be explained by considering the thermal occupation of the motional modes. While writing this manuscript we became aware of this issue and carried out some detailed experimental and numerical studies beyond the scope of the present manuscript and verified that the drop in the coherence time is indeed due to this effect. Its impact can be understood from the second term of Eq.~\eqref{eq:XYwithACShift}. The AC-Stark shift in this case linearly follows the phonon number, thus imperfect ground-state cooling or motional heating increases the dephasing rates in the system. To improve this coherence, we are currently working on implementing better ground-state cooling methods in our system and designing a new ion trap to reduce the motional heating of the ion string.

In the disorder studies, we use a spin-echo sequence to mitigate slow drifts in the experimental parameters. To do so, an echo pulse, i.e. a $\pi$ rotation about the $x$ axis in the middle of the time evolution is applied. The coherence time measured is $T_{\mathrm{coh}}\sim27(2)$~ms for strong disorders and $T_{\mathrm{coh}}\sim27(4)$~ms  for weak disorders. In the case of pure dephasing, where the prepared state is left to evolve under dephasing noise without interactions, we observe that slow detrimental noise components affect the quality of randomized measurements. To overcome this, we again use a spin-echo sequence. The measured coherence time is $T_\mathrm{coh}\sim45(4)$~ms. 

In the numerical simulations shown in the results, we have considered a global dephasing mechanism with the experimentally measured rates. The corresponding jump operator is $\sqrt{\Gamma} \sum_i \sigma_i^z$, where $\Gamma = 1/(2T_{\mathrm{coh}})$. The spontaneous decay rate and depolarization noise rate reported elsewhere~\cite{Joshi2022} are negligible compared to these rates and hence they are not the leading error sources for the current studies. To counteract slow drifts of the experimental parameters, the system is calibrated every few minutes for changes in clock transition, laser pulse area fluctuation, and shift in the ion position.

The impact of decoherence in observing the QMPE is further studied via numerical simulations. The results are displayed in Fig.~\ref{fig:cohLoss} while considering various values of global dephasing rates $\Gamma$. We notice that features like the crossing of EA curves and observation of the QMPE remain unaffected. However, dephasing decreases EA for all tilt angles. The revival of EA is noticeably damped down by dephasing (see the peak at around $t=7.5$ ms). Nonetheless, we can conclude that for qualitative purposes the QMPE is robust under global dephasing.

\subsection{Implementation of disorder in the experiment}
\label{appen:disorders}
\begin{suppfig}
[h!]
    \centering
\includegraphics[width=1\linewidth]{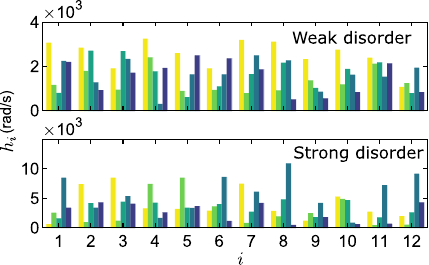}
\caption{Disorder strength $h_i$ versus the ion index for the weak and strong disorder realizations employed in the experiments. Each color corresponds to one of the 5 sets of different realizations of disorders considered.}
\label{fig:disorderrealized}
\end{suppfig}
For the disorder studies presented in the main manuscript, the XY interaction is combined with site-dependent transverse field terms. This disorder Hamiltonian reads as
\begin{equation}
H =\sum_{i>j}\frac{J_0}{2|i-j|^\alpha} \left(\sigma^x_i \sigma^x_j + \sigma^y_j \sigma^y_i\right)+ \sum_i h_i \sigma^z_i~, 
\label{eq:disorderH}
\end{equation}
where $h_i$ indicates transverse disorder strength. In our experimental system, these disorders terms are realized by shining multiple laser beams generated by an acousto-optic deflector (AOD) on individual ions. The AOD deflects multiple laser beams designated to individual ions with the desired light intensity such that site-dependent AC-Stark shifts are induced. We sample $5$ sets of weak and strong disorders randomly chosen from a uniform distribution function and apply them to the ion string. In Fig.~\ref{fig:disorderrealized}, we show the measured disorders for both the strong and weak disorder cases.

\subsection{Measurements}
\label{sec:measurements}
The measurement protocol is shown in Fig.~1a of the main manuscript. We begin by preparing a  ferromagnetic product state \mbox{$\rho^{}_\mathrm{ferro}=(\ket{\downarrow}\bra{\downarrow})^{\otimes N} \equiv (\ket{1}\bra{1})^{\otimes N}$}. This state is $U(1)$ symmetric under rotations about the $z$ axis. Next, we create initial states that explicitly break the $U(1)$ symmetry by rotating the state $\rho^{}_{\mathrm{ferro}}$ about the $-y$ axis using the operator $R_{-y}(\theta)=\exp(i \theta S_y/2)$ with $S_y=\sum_i\sigma^y_i$. This allows us to create the symmetry breaking initial states  $\rho(\theta)_{\mathrm{init}}=R_{-y}(\theta)\rho^{}_{\mathrm{ferro}}R_{y}(\theta)$, where the tilt angle  $\theta$ controls the extent of broken symmetry. This state is quenched with Hamiltonians of interest [Eqs.~2, 3 of the main manuscript] for a duration $t$, leading to the state $\rho(\theta, t)$ that is finally measured in random bases (detailed below). The measurements are performed for each tilt angle $\theta$ and time $t$ independently, thus, hereon we drop the parenthesis to be concise. 

The randomized measurement (RM) framework \cite{elben_review, cieslinski2023analysing} allows us to estimate desired quantities in a state-agnostic manner with a reduced measurement budget compared to the quantum state tomography~\cite{Haah2017}. This toolbox has allowed probing a large range of interesting quantum properties such as entanglement entropies~\cite{Brydges2019,Rath2021,Satzinger2021,Hoke2023}, negativities~\cite{Zhou2020,Elben2020b,Neven2021}, bound entanglement~\cite{Imai2021,zhang2023experimental}, quantum Fisher information~\cite{Rath_QFI_2021,vitale2023estimation} and many-body quantum chaos~\cite{LKJ_2022_chaos}.
The RM protocol requires applying unitaries $U= \bigotimes_{i = 1}^N U_i$ to the time-evolved states, where each single qubit random rotation $U_i$ is sampled from at least a 2-design. Then, the rotated state $U\rho U^\dag$ is projected in the $z$ basis $\ket{\mathbf{s}} = \ket{s_1, \dots, s_N}$ with $s_i \in (0,\,1)$. Note that, in the experiment, we construct local random unitaries $U_i$ by combining global $x$ and $y$ rotations with single qubit local $z$ rotations whose angles are sampled randomly to obtain a set of unitaries from the circularly uniform ensemble (CUE)~\cite{Mezzadri2006}. For the $x$ and $y$ global rotations, we use an elliptically shaped laser beam coupling to all ions simultaneously and driving the resonant optical transition between spin states. A tightly focused laser beam is used to perform local $z$ rotations, by inducing an AC-Stark shift of variable strength on individual ions. Quantum projective measurements in the $z$ basis are obtained via site-resolved fluorescence detection.  

The protocol described above is repeated for $N_U$ different random unitaries and recording $N_M$ projective measurements for each random unitary. The experimental data thus obtained comprises of random unitaries $U^{(r)}$ and the respective recorded bitstring measurements  $\mathbf{s}^{(r,m)} = \left( s_1^{(r,m)}, \dots, s_N^{(r,m)}\right)$ with $r = 1, \dots, N_U$ and $m = 1, \dots, N_M$ that are stored in a classical device for post-processing. From the collected RM dataset, we can construct operators known as \textit{classical shadows} of the prepared state $\rho$ in the post-processing stage.
This operator restricted to a given subsystem $A$ can be defined as~\cite{Huang2020, elben_review}
\begin{equation}
    \hat{\rho}^{(r,m)}_{A} = \bigotimes_{i \in A} 3 \,{U_i^{(r)}}^\dag \ketbra{s_i^{(r,m)}}{s_i^{(r,m)}} U_i^{(r)} - \mathbb{I}_2~. \label{eq:shadow_A}
\end{equation}
These classical shadows form an unbiased estimator of the underlying density matrix of interest, that is, the average over the applied unitaries and the measurement outcomes gives $\mathbb{E}[\hat{\rho}^{(r,m)}_A] = \rho_A$.

\subsection{Estimation  of the EA}
\label{sec:EARandomized}
The shadows defined in Eq.~\eqref{eq:shadow_A} can be used to extract arbitrary multi-copy functionals of the form  \mbox{$f_n = \tr(O_A^{(n)} \rho_A^{\otimes n})$} defined on the subsystem $A$ using the U-statistics estimator~\cite{Hoeffding1992}:
\begin{equation}
     \hat{f}_n = \frac{1}{n!} \binom{N_U}{n}^{-1} \sum_{r_1\ne \dots \ne r_n} \tr\left(O_A^{(n)} \bigotimes_{j = 1}^n \hat{\rho}_A^{(r_j)}\right)~, \label{eq:ustat}
\end{equation}
which is an unbiased estimator of $f_n$, i.e $\mathbb{E}[\hat{f}_n] = f_n$.
We also define here the classical shadow $\hat{\rho}_A^{(r)} = \mathbb{E}_{N_M}[\hat{\rho}_A^{(r,m)}]$, constructed by averaging over $N_M$ measured bitstrings for an applied unitary $U^{(r)}$.
Additionally, the classical shadow formalism allows one to estimate interesting symmetry-resolved quantum properties of the underlying quantum state as shown in the Refs.~\cite{Neven2021,Vitale2022,Rath2023OE}.
In this work, in order to obtain an estimator $\widehat{\rm EA}$ of Eq.~1 (main manuscript) from the RM dataset, we construct unbiased estimators of each of the individual trace moment term in Eq.~1 (main manuscript) following Refs.~\cite{elben_review, Vitale2022}. This is written as,
\begin{equation}
    \widehat{\Delta S_A} = \log \frac{\sum_{r_1 \ne r_2} \tr\left(\hat{\rho}_A^{(r_1)}\hat{\rho}_A^{(r_2)}\right)}{\sum_{r_1 \ne r_2}\tr\left(\hat{\rho}_{A,Q}^{(r_1)}\,\hat{\rho}_{A,Q}^{(r_2)}\right)}~,
\end{equation}
where $\hat{\rho}_{A,Q}^{(r)}$ is the shadow for the symmetrized counterpart of $\hat{\rho}_A^{(r)}$ given by \mbox{$\hat{\rho}_{A,Q}^{(r)} = \sum_{q \in \mathbb{Z}} \Pi_q \hat{\rho}_A^{(r)} \Pi_q$} that satisfies $\mathbb{E}[\hat{\rho}^{(r)}_{A, Q}] = \rho_{A, Q}$. Rigorous performance analysis of statistical errors from a finite number of experimental runs $N_U\times  N_M$ in the estimation of such trace moments have been studied in various prior works~\cite{Huang2020,Elben2020b,Neven2021,Rath_QFI_2021}.

\subsection{Frobenius distance}
\label{sec:Frobdis}
For generic large quantum systems evolving unitarily, the reduced density matrix describing the subsystem $A$ is expected to relax, at long times, to a Gibbs distribution~\cite{Deutsch1991, Srednicki1994, rigol2008}. The eigenstate thermalization hypothesis~\cite{Deutsch1991, Srednicki1994} predicts that such a thermal state is equivalent to the reduced density matrix of the diagonal ensemble~\cite{rigol2007, rigol2008}. The diagonal ensemble is obtained by projecting the initial density matrix onto the eigenbasis $\{\ket{E_k}\}$ of the quenching Hamiltonian, that is
\begin{eqnarray}
\rho^{\mathrm{DE}}(\theta)&=& \sum_k \bra{E_k}\rho(\theta, t=0)\ket{E_k} \ketbra{E_k}{E_k}~,
\label{eq:diag_dm}
\end{eqnarray}
where $\rho(\theta, t=0)$ is the initial state. The partial trace of this state $\rho_A^{\mathrm{DE}}(\theta)=\tr_{\bar A}(\rho^{\mathrm{DE}}(\theta))$ gives the average behavior of local operators at long times. For each tilt angle $\theta$, we are interested in the Frobenius distance between the post-quench state of subsystem $A$ at a given time and the corresponding diagonal ensemble i.e., the distance FD$[\rho_A(\theta,t), \rho_A^{\mathrm{DE}}(\theta)]$, defined as~\cite{fagotti2013reduced}, 
\begin{equation}
    \mathrm{FD}(\rho_A, \rho_A^{\mathrm{DE}})= \sqrt{1-\mathrm{min}\left(1, \frac{2\mathrm{Tr}[\rho_A\rho_A^\mathrm{DE}]}{\mathrm{Tr}[\rho_A^2+(\rho_A^\mathrm{DE})^2]}\right)},
    \label{eq:FrobDist}
\end{equation}
where we have dropped the angle $\theta$ and time $t$ for brevity. Here, we have put a bound on the second term to bypass the unphysical Fidelities. In the limit of infinitely many measurements, the second term should always be upper bound by 1. In the main text, we utilized this distance to analyze the thermalization of the reduced density matrix after the XY Hamiltonian quench (see Fig.~4). 

The rich RM dataset obtained using the protocol described above equally allows us to extract the Frobenius distance. Namely, we can estimate unbiased estimators of all the individual terms in the expression of the Frobenius distance~\eqref{eq:FrobDist} from the \textit{same} dataset collected during the experimental execution. Thus the explicit expression of the estimator $\widehat{\mathrm{FD}}$ of the Frobenius distance can be written as
\begin{equation}
    \widehat{\mathrm{FD}}= \left(1 - \frac{\frac{2}{N_U}\sum_{r_1}\tr[\hat{\rho}_A^{(r_1)}\rho_A^{\mathrm{DE}}]}{ \beta^{-1} \sum_{r_1 \ne r_2} \tr[\hat{\rho}_A^{(r_1)} \hat{\rho}_A^{(r_2)}]+\tr[(\rho_A^{\mathrm{DE}})^2]}\right)^{1/2}
\end{equation}
with $\beta = N_U(N_U-1)$.

\subsection{Estimation of error bars} For randomized measurements, experiments are carried out for $N_U$ random unitaries and $N_M$ number of projective measurements for each random unitary. In the main text, the time evolution with the XY interaction (Fig.~1b) is done with $N_U = 500$ and $N_M = 30$. For the inset of Fig.~1b, we sample $N_U = 200$ and $N_M = 100$. The experiments in the presence of disorder (results in Fig.~2) and pure dephasing (results in Fig.~3) are done with $N_U = 300$ and $N_M = 30$. We used a total of 5 disorder sets for each $\theta$ in the disorder studies.

In the main text, for the disorder case, the error bars are estimated from the standard deviation of the mean over disorder realization. Elsewhere, the error bars are estimated via the Jackknife method applied on the random unitaries that are used to estimate the averaged EA \cite{wu1986jackknife, elben_review}. Wherever the subsystem average data are presented, we employ the Jackknife resampling method by dropping one random unitary at a time and subsequently estimating the variance of the subsystem averaged EA. The subsequent error bars of the mean are presented in the figures.

\subsection{Finite size effects: $4$-qubit experiment}
\label{appen:4_qubit_EA}
\begin{suppfig}[h!]
\includegraphics[width=1\linewidth]{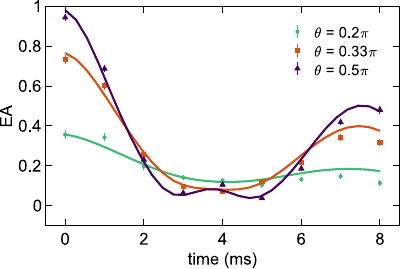}
\caption{The EA after a quench with the XY Hamiltonian on a system of $N=4$ qubits. The averaged EA over all possible subsystems of size $N_A=2$  is plotted for three tilt angles. Symbols are the experimental data and curves correspond to numerical simulations. The subsystem restores the symmetry and the QMPE is observed. However, a revival of the EA is seen due to the finite-size effects.}
\label{fig:4_qubit_ea}
\end{suppfig}

To explore the finite-size effects within the experiment's coherence time, we measure the EA for a $4$-qubit system and present the results of the EA for $N_A=2$ subsystems. Data are plotted in Fig.~\ref{fig:4_qubit_ea}. Here, symbols correspond to experimental data, and solid lines correspond to numerical simulations in the presence of global dephasing. It is important to note that QMPE is present even for the smallest system that we probed in the experiment. However, a revival of the EA is observed due to finite-size effects. In fact, in this case, thermalization is hindered and thus the subsystem oscillates between the asymmetric and symmetric states.
\bibliography{BibArxivV2}

\end{document}